\documentclass[journal]{IEEEtran}
\usepackage{cite}
\usepackage{amsmath}
\usepackage{geometry}
\usepackage{hyperref}
\usepackage{siunitx}
\usepackage{graphicx} 
\usepackage[table]{xcolor}
\usepackage{booktabs}
\usepackage{multirow}
\usepackage{makecell}
\usepackage{tikz}
\usetikzlibrary{shapes.geometric, arrows.meta, positioning}
    \tikzstyle{block} = [rectangle, rounded corners, minimum width=6.5cm, minimum height=1.4cm, text centered, draw=black, fill=blue!10, align=center]
    \tikzstyle{subblock} = [rectangle, rounded corners, text width=5.5cm, text centered, draw=black, fill=gray!20, font=\small]
    \tikzstyle{repeatblock} = [rectangle, rounded corners, text width=5.5cm, text centered, draw=black, fill=gray!20, font=\footnotesize]
    \tikzstyle{arrow} = [thick, ->, >=Stealth]

\begin{document}

\title{Electromagnetic Noise Characterization and Suppression in Low-Field MRI Systems}

\author{\IEEEauthorblockN{
		Teresa\,Guallart-Naval\IEEEauthorrefmark{1},
		Jos\'e~M.~Algar\'in\IEEEauthorrefmark{1} and
		Joseba\,Alonso\IEEEauthorrefmark{1}}

		\IEEEauthorblockA{\IEEEauthorrefmark{1}MRILab, Institute for Molecular Imaging and Instrumentation (i3M), Spanish National Research Council (CSIC) and Universitat Polit\`ecnica de Val\`encia (UPV), 46022 Valencia, Spain}\\

\thanks{Corresponding author: J. Alonso (joseba.alonso@i3m.upv.es).}}

\maketitle

\begin{abstract}
	
\textbf{Purpose:} Our goal is to develop and validate a practical protocol that guides users in identifying and suppressing electromagnetic noise in low-field MRI systems, enabling operation near the thermal noise limit.

\textbf{Methods:} We present a systematic, stepwise methodology that includes diagnostic measurements, hardware isolation strategies, and good practices for cabling and shielding. Each step is validated with corresponding noise measurements under increasingly complex system configurations, both unloaded and with a human subject present.

\textbf{Results:} Noise levels were monitored through the incremental assembly of a low-field MRI system, revealing key sources of EMI and quantifying their impact. Final configurations achieved noise within 1.5× the theoretical thermal bound with a subject in the scanner. Image reconstructions illustrate the direct relationship between system noise and image quality.

\textbf{Conclusion:} The proposed protocol enables low-field MRI systems to operate close to fundamental noise limits in realistic conditions. The framework also provides actionable guidance for the integration of additional system components, such as gradient drivers and automatic tuning networks, without compromising SNR.
\end{abstract}

\section{Introduction}

Low-field MRI systems, operating in the \SIrange{1}{10}{MHz} range, are gaining renewed attention due to their cost-effectiveness~\cite{Webb2023}, portability~\cite{Algarin2024}, inherent safety advantages~\cite{Shellock2024}, and DIY compatibility (even if nontrivial in practice)~\cite{Huang2025}. These features make them well-suited for point-of-care imaging, low-resource settings, and pediatric applications. However, a major challenge at low magnetic fields is the limited signal-to-noise ratio (SNR)~\cite{Webb2023b,Ayde2024}.

At MHz-range Larmor frequencies, the dominant source of intrinsic noise is thermal (Johnson-Nyquist) noise due to the RF coil resistance~\cite{Webb2023b}. In principle, the receive chain can approach the thermal limit if all other noise contributions—such as electromagnetic interference (EMI) from cabling, active electronics, gradient drivers, and the imaging subject—are sufficiently suppressed. Despite the fundamental nature of the thermal noise limit, it is often overlooked in the design and evaluation of low-field platforms. Studies reporting noise measurements benchmarked directly against the theoretical baseline set by a \SI{50}{\ohm} resistor are notoriously scarce \cite{Lena2025}. Numerous prior works have addressed EMI suppression or active noise cancellation (see e.g. \cite{Liu2021,Srinivas2022,Parsa2023,Zhao2024}). However, despite operating in signal space, they typically quantify performance in terms of relative noise reduction—comparing noise levels before and after applying the suppression technique directly in image space—without referencing the absolute thermal noise floor. This leaves the reader uncertain about the true effectiveness of the system. The adoption of standardized metrics should arguably help unify performance reporting and facilitate the transition of low-field technologies into clinical and translational settings.

In this work, we present a practical, step-by-step methodology for characterizing and suppressing electromagnetic noise in low-field MRI systems. Our approach is based on progressive system assembly, diagnostic noise measurements, and rigorous EMI mitigation strategies. The protocol is designed to help researchers identify dominant noise sources and guide them in achieving operation close to the fundamental thermal limit. To validate the methodology, we present measurements of system noise as each component is added, and we demonstrate the direct impact of noise suppression on image quality through a series of representative scans. The methodology is broadly applicable to low-field systems of various designs, particularly those operating in challenging electromagnetic environments or developing custom low-field MRI systems without industrial shielding and integration standards.

\section{Theory}
\label{sec:theory}

Johnson-Nyquist noise arises from the random motion of charge carriers in resistive elements~\cite{Johnson1928}. The root-mean-square (RMS) voltage noise generated by a resistor $R$ over a bandwidth $\Delta f$ at temperature $T$ is given by
\begin{equation}
	v_{\mathrm{n}}^2 = 4 k_\text{B} T R \Delta f,
\end{equation}
where $k_\text{B} \approx \SI{1.38e-23}{J/K}$ is Boltzmann’s constant. For a coil impedance matched to $R = \SI{50}{\ohm}$ and operating near room temperature ($T \approx \SI{300}{K}$), the corresponding voltage noise spectral density is
$\approx \SI{0.9}{nV}/\sqrt{\mathrm{Hz}}$. This value serves as a universal reference point for noise performance in low-field systems. If the RF coil is connected to a \SI{50}{\ohm} input preamplifier this value halves to
\begin{equation*} 
	v_{\mathrm{n}}/\sqrt{\mathrm{Hz}} \approx \SI{0.45}{nV/\sqrt{Hz}}.
\end{equation*}

In practice, the receive chain includes a low-noise amplifier (LNA), whose output voltage is expected to be
\begin{equation}\label{eq:baseline}
	v_{\mathrm{out}} = G\cdot \sqrt {k_\text{B} T R \Delta f},
\end{equation}
where $G$ is the linear transducer gain of the amplifier\footnote{The transducer gain is defined as the power delivered to a \SI{50}{\ohm} load divided by the power available at the amplifier input. Manufacturers typically specify the power gain in dB, $G = 10^{G_\text{dB}/20}$.}. Equation~(\ref{eq:baseline}) defines the theoretical baseline against which system noise measurements are to be compared. Any observed noise significantly above this threshold is indicative of external EMI or design flaws in cabling, grounding, or shielding\footnote{In fact, the LNA also introduces noise, typically characterized by the noise factor $\text{NF} = 10^{\text{NF}_\text{dB}/20}$, so Eq.\,(\ref{eq:baseline}) becomes $v_{\mathrm{out}} = G \cdot \text{NF} \cdot \sqrt{k_\text{B} T R \Delta f}$. A high-performance LNA might have $\mathrm{NF}_\text{dB} = \SI{1}{dB}$, corresponding to $\text{NF} \approx 1.12$, resulting in a modest increase in the noise floor.}.

This reference model assumes the RF coil behaves like an ideal \SI{50}{\ohm} resistor—an approximation that holds only when the coil is both well-tuned and impedance-matched, and only at the center frequency. Likewise, amplifier specifications (gain and noise factor) are typically provided for \SI{50}{\ohm} input/output impedances. As a result, the thermal baseline is most accurate for narrowband measurements taken well within the resonance linewidth of the RF coil. In practice, for coils that are properly tuned and matched, noise measurements with bandwidths below \SI{20}{kHz} are often experimentally indistinguishable from those obtained using a true \SI{50}{\ohm} terminator. Nonetheless, using wider acquisition bandwidths—such as those compatible with typical imaging sequences—can be advantageous for identifying discrete EMI sources, which often manifest as prominent spectral spikes, provided they fall within the measurement range.

\section{Methods}

\subsection{Experimental setup}
\label{sec:apparatus}

\begin{figure*}[htbp]
	\centering
	\includegraphics[width=1\textwidth]{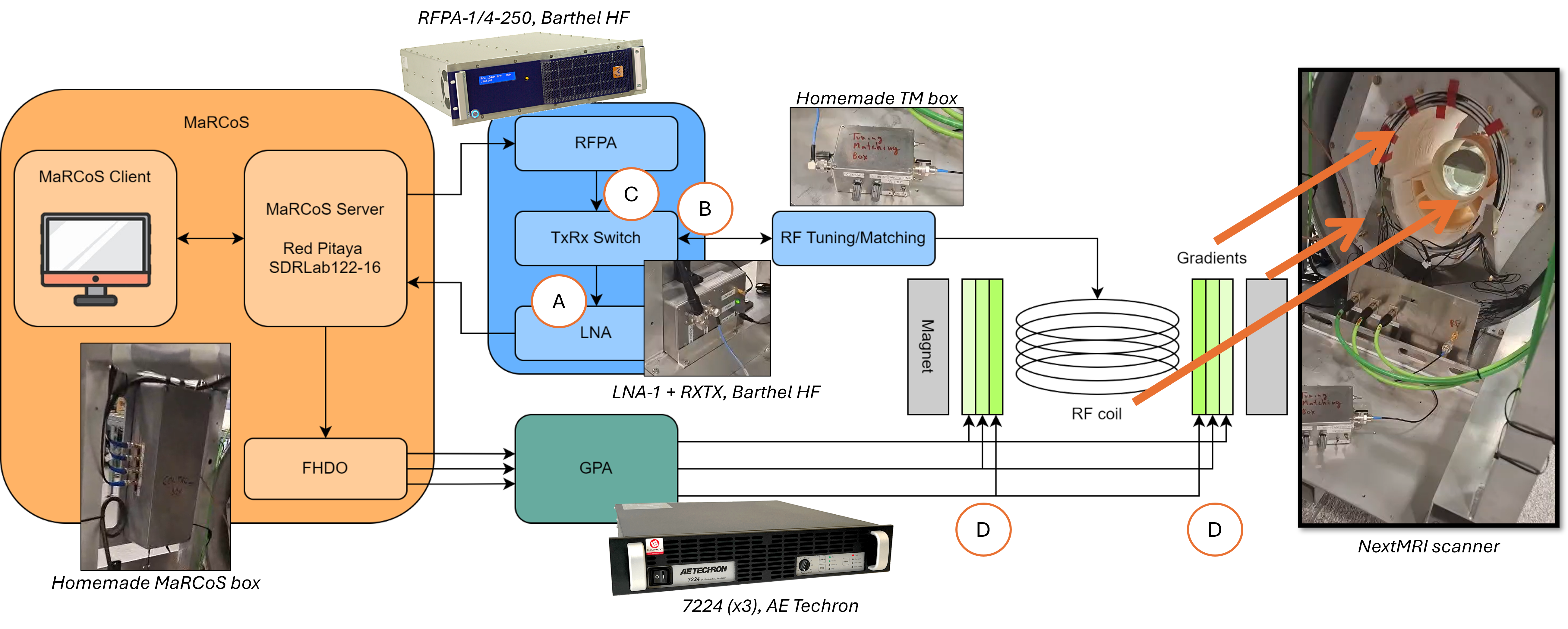}
	\caption{Diagram of the main components in a low-field MRI scanner employing MaRCoS, including details and photographs of these components in the NextMRI system employed for this work. Figure adapted from Ref.~\cite{Huang2025}.}
	\label{fig:setup}
\end{figure*}

All measurements and imaging in this work are performed on a custom low-field MRI scanner (Fig.~\ref{fig:setup}) developed in Valencia as part of the NextMRI project~\cite{Galve2024,Pallas2025}. The scanner includes an elliptical permanent magnet in a Halbach configuration. Gradient coils are made from water-jetted copper plates shaped using elliptical molds and secured to 3D-printed PLA supports. The gradients are driven from three independent commercial gradient power amplifiers (GPA) capable of delivering up to \SI{50}{A} and \SI{158}{V} per channel (AE Techron 7224). The single RF transmit and receive coil is a spiral head coil based on a design presented in Ref.~\cite{Sarracanie2015}, optimized for whole-brain coverage at low field strengths. The RFPA, LNA and TxRx switch are commercial modules (Barthel HF) with \SI{50}{\ohm} input impedance. The specified LNA gain and noise factor are $\approx\SI{45}{dB}$ ($G\approx178$) and $\approx\SI{1}{dB}$ ($\mathrm{NF}\approx 1.12$), respectively.

The entire assembly—including the magnet, shimming unit, gradient coils, RF electronics, and control hardware—is housed in an aluminum structure of $68 \times 95 \times \SI{140}{cm^3}$ (width $\times$ height $\times$ length). This structure is mounted on a mobile platform and the system is fully self-contained and operates from a standard single-phase power outlet.

Control of the scanner is managed by the open-source MaRCoS (Magnetic Resonance Control System) framework~\cite{Negnevitsky2023,GuallartNaval2022b}. MaRCoS combines dedicated hardware, firmware, and software for synchronized pulse sequence execution and data acquisition. User interaction is handled via the MaRGE interface (MaRCoS Graphical Environment)~\cite{Algarin2024b}, which provides control of the scanner through a tabbed graphical layout. MaRGE includes numerous pulse sequences by default, including system calibrations and noise measurements. It is also compatible with Pulseq \cite{Layton2017} and supports image export to DICOM, protocol management, and basic image reconstruction tools.

\subsection{Noise measurements}
\label{sec:measurements}

All noise measurements described in this work are performed using a consistent acquisition procedure. The basic setup consists of the receive chain—including the amplification stage (typically a low-noise amplifier followed by optional secondary gain stages), an analog-to-digital converter (ADC), control electronics, and a control computer running the user interface. In our case, a dedicated noise acquisition sequence is programmed in MaRGE to open the ADC for a fixed acquisition duration and return the recorded signal for analysis. The data follows from the internal signal pre-processing configured in the control system, such as oversampling, decimation, and filtering \cite{Negnevitsky2023}. This procedure can be extrapolated to any other control system. If this does not count on a dedicated noise measurement sequence, the same information can be obtained from a simple FID with zero amplitude on the RF excitation pulse.

Throughout this work, a typical measurement uses an acquisition bandwidth of \SI{50}{kHz}, centered at the system’s operating frequency. Acquisition durations are chosen to cover at least one full \SI{50}{Hz} mains cycle (e.g., \SI{50}{ms}), ensuring visibility of periodic EMI bursts from power-line sources. The noise amplitude is estimated from the root-mean-square value of either the real (in-phase) or imaginary (quadrature) channel. The measured RMS voltage needs to be doubled before comparing against the value of $v_\text{out}$ determined from Eq.~(\ref{eq:baseline}), since half the voltage is lost to the term that oscillates at roughly twice the demodulation frequency (and is filtered away). In the employed MaRGE release, this factor of 2 is included by default. As a final note, sizable relative fluctuations (often $>10\,\%$) are to be expected from measurement to measurement due to noise sources and EMI pickup changing slowly compared to an acquisition window (tens of ms), but significantly from shot to shot.

\subsection{Image acquisitions}
\label{sec:imaging}

In addition to noise measurements, we perform representative image acquisitions to illustrate the impact of electromagnetic noise on image quality. For this purpose, we employ 3D RARE (Rapid Acquisition with Relaxation Enhancement) pulse sequences with T1-weighting. The sequence parameters were: repetition time \SI{600}{ms}, echo time \SI{32}{ms}, echo train length $4$, spatial resolution $1.8\times2.1\times\SI{5.3}{mm^3}$, field of view $200\times230\times\SI{160}{mm^3}$, bandwidth \SI{73.4}{kHz}, partial Fourier fraction $0.85$, and total scan time \SI{7}{min}.

\subsection{Noise characterization and suppression protocol}
\label{sec:protocol}

\begin{figure}
	\centering
	\begin{tikzpicture}[node distance=2.3cm]
		
		\node (step1) [block] {
			\begin{minipage}{6cm}
				Step 1: Establish a baseline (Eq.~(\ref{eq:baseline}))
			\end{minipage}
		};
		
		\node (step2) [block, below= 0.3cm of step1] {
			Step 2: Measure noise with a 50~$\Omega$ resistor
		};
		
		\node (substeps) [subblock, below=0.05cm of step2] {
			I. Minimal Rx chain\\
			II. Insert the Tx/Rx switch\\
			III. Connect the transmit (Tx) chain\\
			IV. Power on the RFPA\\
			V. Connect the gradient cables\\
			VI. Power on the GPA\\
			VII. Integrate auxiliary subsystems
		};
		
		\node (step3) [block, below=3.4cm of step2] {
			Step 3: Add RF coil and tune/match
		};
		\node (step3sub) [repeatblock, below=0.05cm of step3] {
			Repeat stages I–VII with RF coil input
		};
		
		\node (step4) [block, below=0.9cm of step3] {
			Step 4: \emph{In vivo} noise evaluation
		};
		\node (step4sub) [repeatblock, below=0.05cm of step4] {
			Repeat stages I–VII with subject in scanner
		};
		
		\draw [arrow] (step1) -- (step2);
		\draw [arrow] (substeps.south) -- (step3);
		\draw [arrow] (step3sub.south) -- (step4);
		
	\end{tikzpicture}
	\caption{Four-step protocol for electromagnetic noise characterization and suppression in low-field MRI systems. Steps 2–4 involve repeated measurements through stages I–VII to assess noise contributions at each level of system assembly.}
	\label{fig:protocol_flowchart}
\end{figure}

\subsubsection*{Step 1: Establish a baseline}

First, we followed the framework in Section~\ref{sec:theory} to compute the expected thermal noise floor for the receive chain (Eq.~(\ref{eq:baseline})). This serves as the benchmark for all subsequent noise measurements.

\subsubsection*{Step 2: Measure noise with a \SI{50}{\ohm} resistor}

\begin{enumerate}
	\item[I] \textbf{Set up a minimal Rx chain.} This consists of LNA, ADC, control electronics, and control computer. Ideally, all other components—including the RF power amplifier (RFPA), transmit/receive (Tx/Rx) switch, gradient drivers, and coil—are physically disconnected from this minimal setup (if possible).
\end{enumerate}

With this minimal setup, we connected a \SI{50}{\ohm} terminator at the LNA input (point \emph{A} in Figure~\ref{fig:setup}), performed noise measurements, and compared them to the baseline established in Step~1. With the resistor, the system should operate within $\sim$1.5$\times$ the theoretical noise floor. If excess noise is detected, at least one of the following actions may be necessary: power the system using batteries or an isolated uninterruptible power supply (UPS), replace RF cables sequentially to rule out shielding failures or broken contacts, swap out active components (starting with the LNA) to identify faulty elements.

We then reconstructed the system incrementally, adding one component at a time and following closely the guidelines in the Appendix. After each addition, we performed new noise measurements. This process enables identification of stages that introduce excess noise. Components should be added in the following order:

\begin{enumerate}
	\item[II] \textbf{Insert the Tx/Rx switch.} This is typically the first element following the RF coil and tuning/matching (TM) circuit. Passive switches with good isolation and shielding are expected to have negligible impact on the noise floor. Connect a \SI{50}{\ohm} terminator at point \emph{B} and \emph{C} in Figure~\ref{fig:setup}. For active switches, set the switch to Rx mode.
	
	\item[III] \textbf{Connect the transmit (Tx) chain (point \emph{C} in Figure~\ref{fig:setup}).} This includes all components responsible for RF pulse generation—such as direct digital synthesizers (DDS), digital-to-analog converters (DACs), and any associated filtering—up to the input of the RF power amplifier (RFPA) and the Tx port of the Tx/Rx switch.
	
	\item[IV] \textbf{Power on the RFPA.} With no RF transmission occurring, the RFPA should remain electromagnetically quiet. Nonetheless, low-frequency noise from its power supply or control logic may couple into the receive path. If the noise level is significantly higher than the thermal reference, the RFPA may require deblanking or the TxRx switch diodes may be faulty.
	
	\item[V] \textbf{Connect the gradient cables (points \emph{D} in Figure~\ref{fig:setup}).} These cables can act as unintended antennas for EMI, particularly if unshielded or routed poorly. Confirm that all cable shielding is intact and that grounding is robust.
	
	\item[VI] \textbf{Power on the gradient power amplifier (GPA).} Even when idle, gradient amplifiers may contribute noise. Ensure that EMI suppression measures are effective and that the receive noise remains within acceptable bounds.
	
	\item[VII] \textbf{Integrate auxiliary subsystems.} Connect and activate optional components such as automatic tuning and matching (auto-TM) units, temperature monitors, or active noise cancellation hardware.
\end{enumerate}

If a particular component results in a measurable increase in noise, it is convenient to investigate its power supply, shielding integrity, and grounding (refer to the Appendix for guidance). Whenever possible, decouple its power source from the receive chain and re-test. In some cases, temporary removal or targeted filtering may help isolate and resolve the issue.

\subsubsection*{Step 3: Add RF coil and tune/match}

We replaced the \SI{50}{\ohm} load with the actual RF receive coil and its tuning/matching (TM) circuit. Using a vector network analyzer (VNA), we ensured that the coil is tuned and matched to better than \SI{-20}{dB} return loss at the system’s operating frequency.

We repeated I-VII in Step~2 using the coil input, monitoring the noise level at each stage. For systems employing different RF coils, it is worth following the full procedure first with the smallest one, which will be less prone to noise pickup from the internal shield, and then repeat for the rest in order of increasing size. If an automatic TM unit is planned, one can begin with a passive (manual) TM network for an initial characterization. Auto-TM circuits typically include microcontrollers, relays, voltage regulators, and embedded vector network analyzers~\cite{Bosch2024}—all of which are potential sources of broadband digital noise that couples directly into the receive path. These systems should only be activated once the manual baseline is established.

\subsubsection*{Step 4: \emph{In vivo} noise evaluation}

The final stage of system characterization involved repeating I-VII in Step~2 with a human subject positioned inside the scanner.

\section{Results}

\begin{table}
	\centering
	\caption{Noise levels at different stages of system assembly (I-VII), normalized to the thermal floor of \SI{18}{\micro V}. The LNA and TxRx switch are part of a single commercial module, which precludes operation without the latter (Row I). Row VII is empty because we are not using any additional equipment.}
	\begin{tabular}{lccc}
		\toprule
		& \textbf{50~$\Omega$} & \textbf{Coil} & \textbf{In vivo} \\
		\midrule
		\textbf{I. Minimal assembly} & - & - & - \\
		\textbf{II. TxRx switch} & 1.26 & 1.32 & 1.32 \\
		\textbf{III. Tx chain conn.} & 1.25 & 1.37 & 1.35 \\
		\textbf{IV. RFPA ON} & 1.23 & 1.31 & 1.33 \\
		\textbf{V. Grads conn.} & 1.25 & 1.46 & 1.56 \\
		\textbf{VI. GPA ON} & 1.24 & 1.53 & 1.51 \\
		\textbf{VII. Other equip.} & - & - & - \\
		\bottomrule
	\end{tabular}
	\label{tab:noise_stages}
\end{table}

The thermal baseline (Step~1) follows directly from Eq.~(\ref{eq:baseline}) and the LNA parameters given in Sec.~\ref{sec:apparatus}, and amounts to $\approx\SI{18}{\micro V}$ in our setup.

Table~\ref{tab:noise_stages} summarizes the noise measurements obtained while following the stepwise procedure described in Sec.~\ref{sec:protocol}. Each row corresponds to a specific stage in the incremental assembly of the MRI system, and measurements are reported under three different loading conditions (columns): with a precision \SI{50}{\ohm} terminator, with the actual RF coil, and with a human subject in the scanner. The values provide a quantitative basis for assessing the impact of each component and configuration on the overall system noise floor.

\def\figsize{3.2}
\begin{table*}[htbp]
	\centering
	\caption{Noise measurements for selected scanner configurations with an unloaded coil. Each scenario has been intentionally recreated to illustrate the impact of specific hardware and grounding practices on electromagnetic noise. For each configuration, the table shows the time-domain trace, power spectrum, and measured RMS noise level normalized against the thermal floor ($\approx\SI{18}{\micro V}$). MaRCoS is the main control box with the ADCs, DACs, and other electronics. TM refers to the box housing the impedance tuning and matching network. The inner shield is a thin copper cylinder between the RF and the gradient coils. The external shield is a thicker aluminum barrel around the magnet. ``Gradients disconnected'' refers to physically removing the connection between the GPA and the gradient coils.}
	\label{tab:time_trace_spectra}
	\begin{tabular}{llccc}
		\toprule
		\textbf{Scanner element} & \textbf{Configuration} & \textbf{Time trace} & \textbf{Power spectrum} & \textbf{Noise level}\\
		
		\midrule
		
		\multirow{2}{*}{RF cables} 
		& \makecell{All shielded} & \includegraphics[width=\figsize cm]{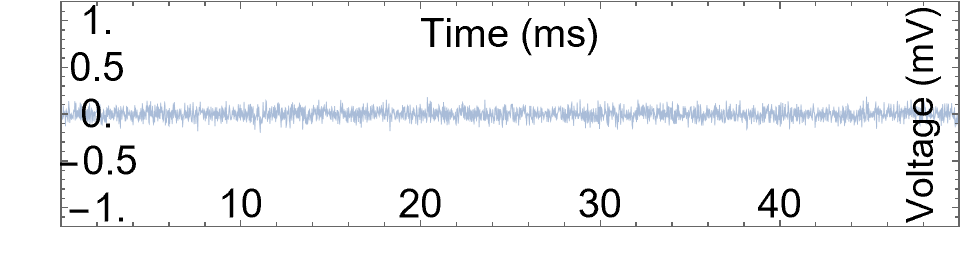} 
		& \includegraphics[width=\figsize cm]{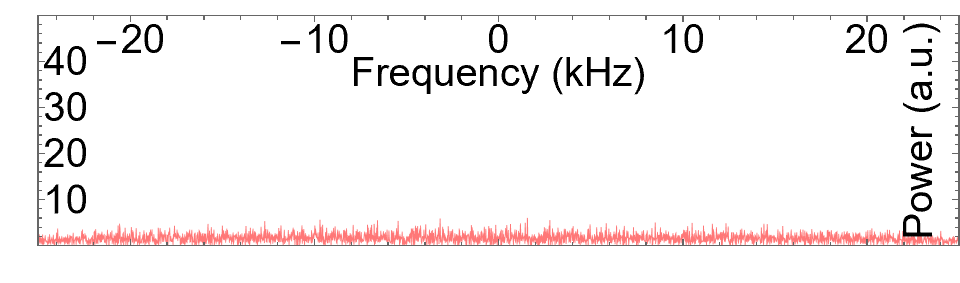} 
		& 1.4$\times$ \\
		& \makecell{One unshielded} & \includegraphics[width=\figsize cm]{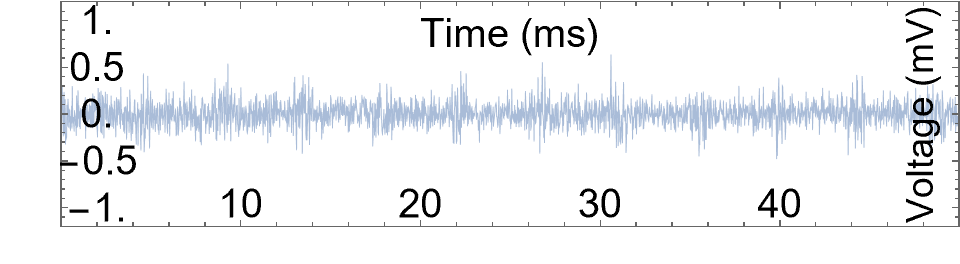} 
		& \includegraphics[width=\figsize cm]{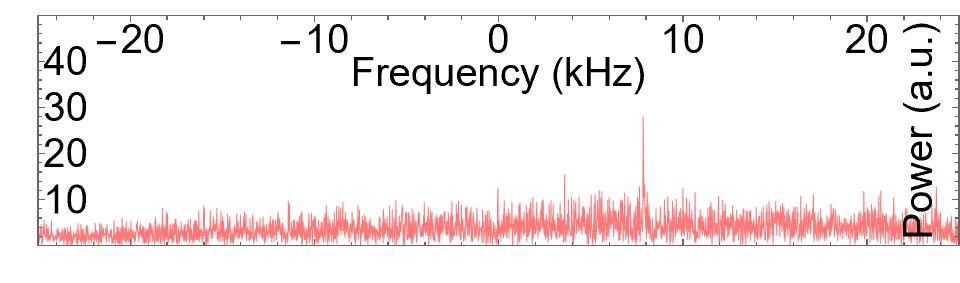} 
		& 3.4$\times$ \\
		
		\midrule
		
		\multirow{2}{*}{Gradient cables} 
		& \makecell{All shielded} & \includegraphics[width=\figsize cm]{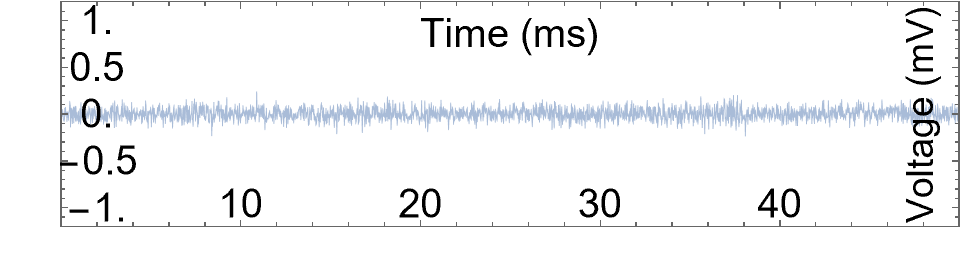} 
		& \includegraphics[width=\figsize cm]{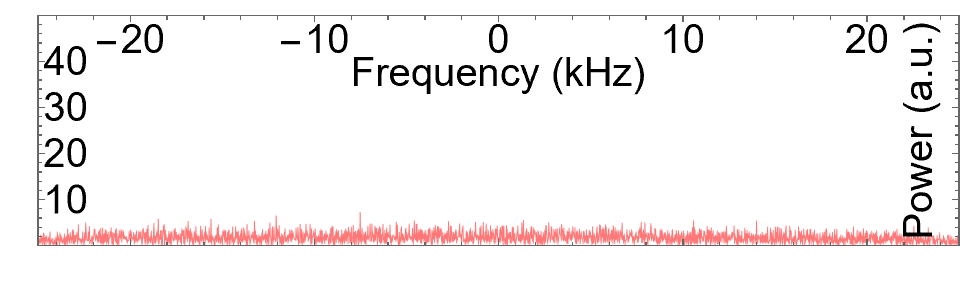} 
		& 1.6$\times$ \\
		& \makecell{One unshielded} & \includegraphics[width=\figsize cm]{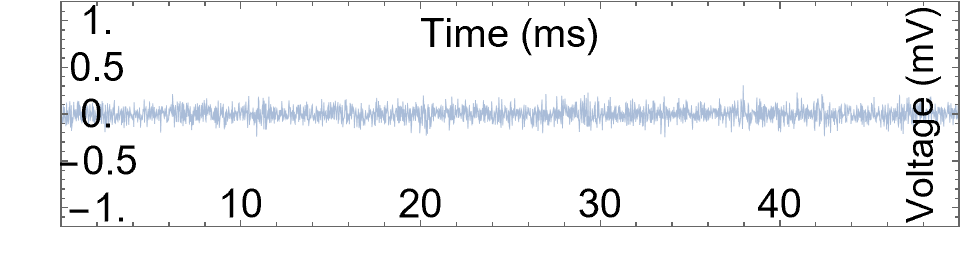} 
		& \includegraphics[width=\figsize cm]{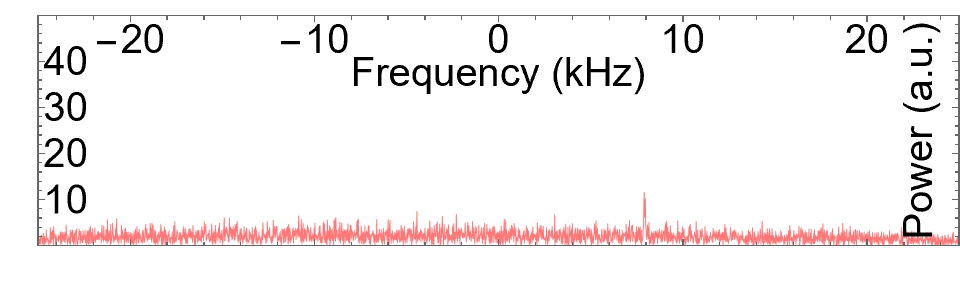} 
		& 1.8$\times$ \\
		
		\midrule
		
		\multirow{4}{*}{Metallic enclosures} 
		& \makecell{All closed} & \includegraphics[width=\figsize cm]{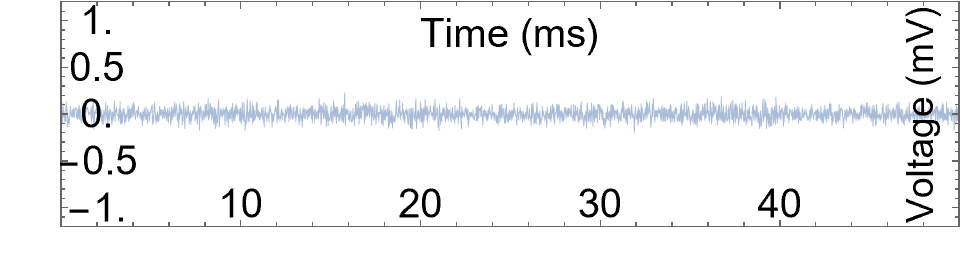} 
		& \includegraphics[width=\figsize cm]{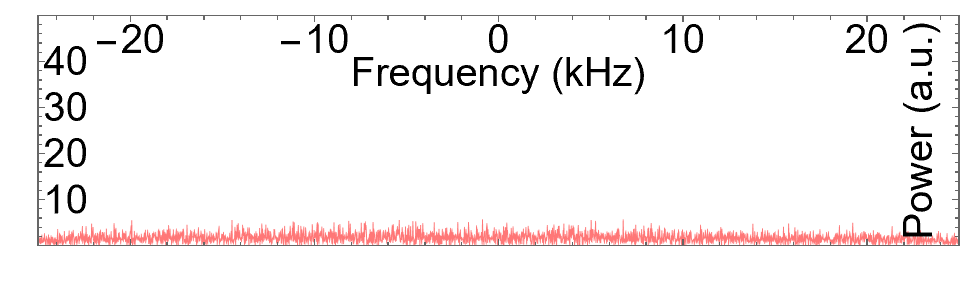} 
		& 1.5$\times$ \\
		& \makecell{MaRCoS open} & \includegraphics[width=\figsize cm]{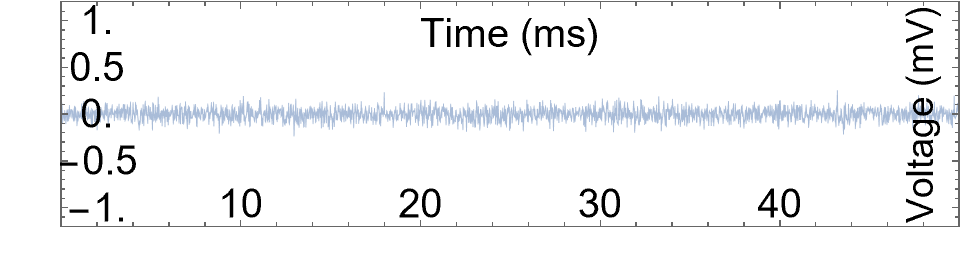} 
		& \includegraphics[width=\figsize cm]{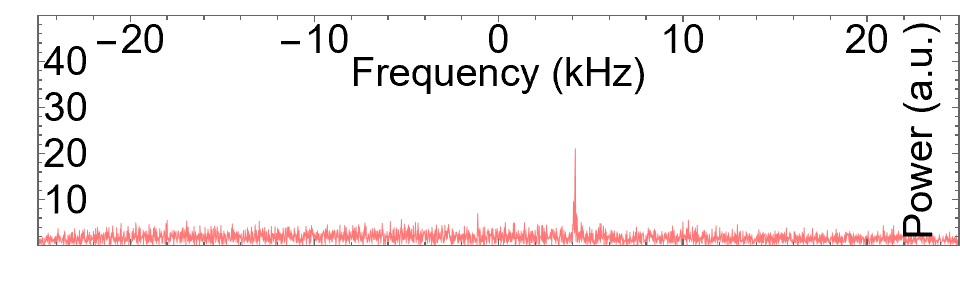} 
		& 1.6$\times$ \\
		& \makecell{TM open} & \includegraphics[width=\figsize cm]{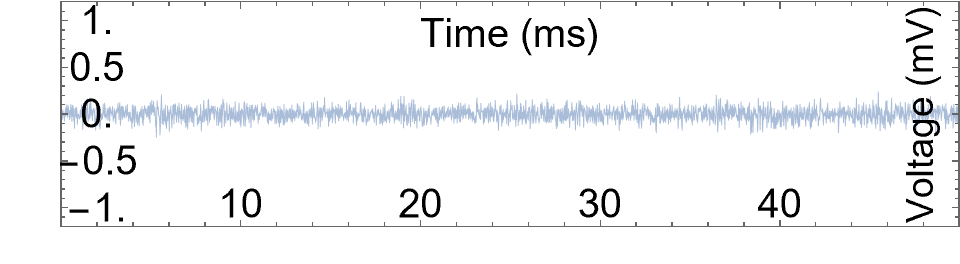} 
		& \includegraphics[width=\figsize cm]{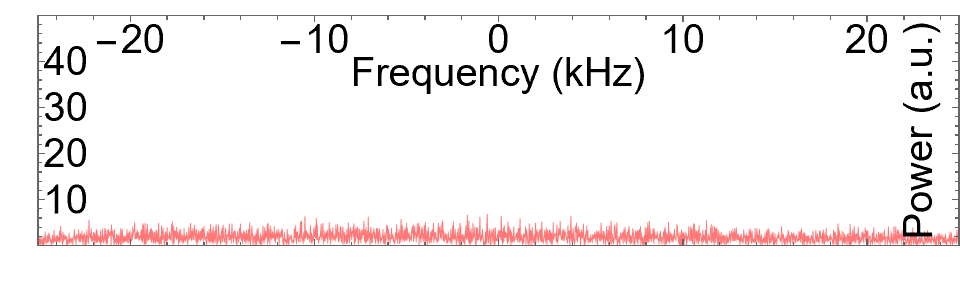} 
		& 1.7$\times$ \\
		& \makecell{MaRCoS and\\ TM open} & \includegraphics[width=\figsize cm]{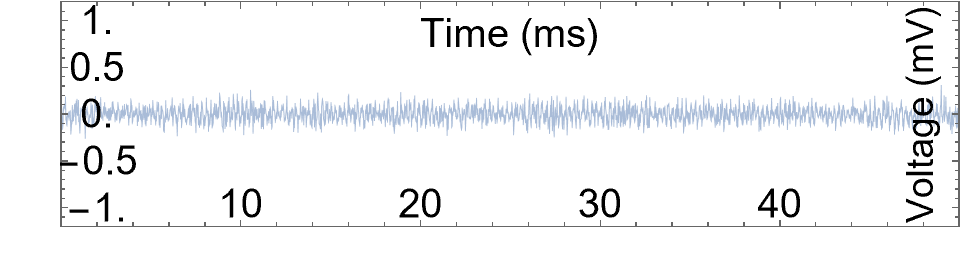} 
		& \includegraphics[width=\figsize cm]{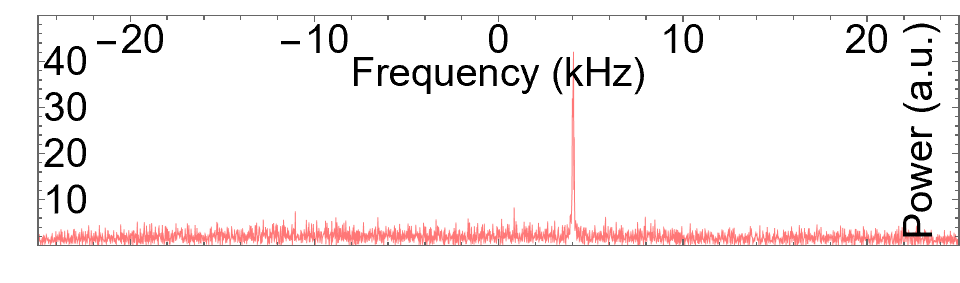} 
		& 2.1$\times$ \\
		
		\midrule
		
		\multirow{5}{*}{Scanner shields} 
		& \makecell{Both grounded,\\ all on} & \includegraphics[width=\figsize cm]{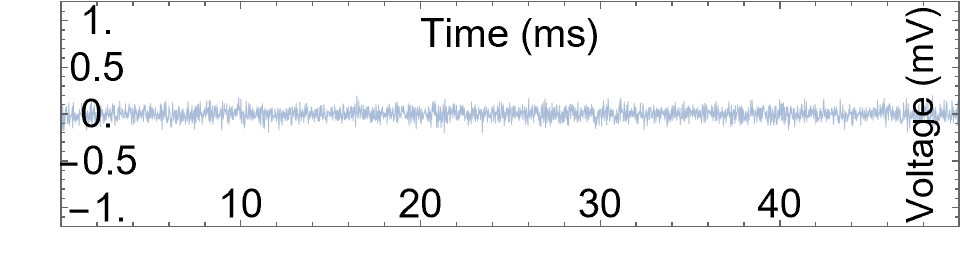} 
		& \includegraphics[width=\figsize cm]{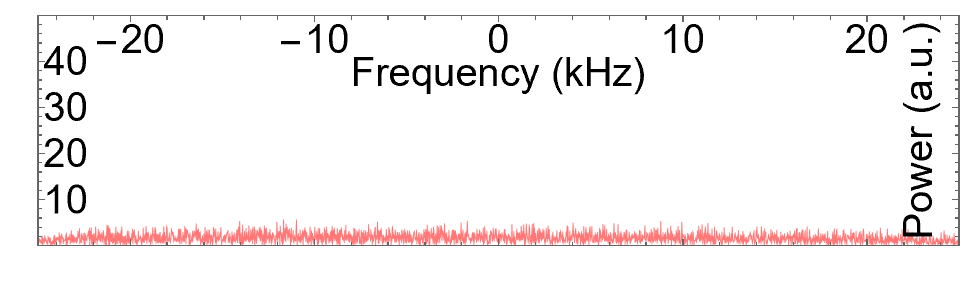} 
		& 1.5$\times$ \\
		& \makecell{Inner floating,\\ gradients disconnected} & \includegraphics[width=\figsize cm]{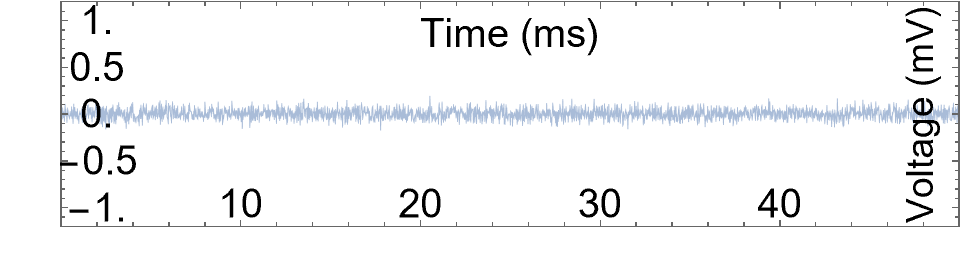} 
		& \includegraphics[width=\figsize cm]{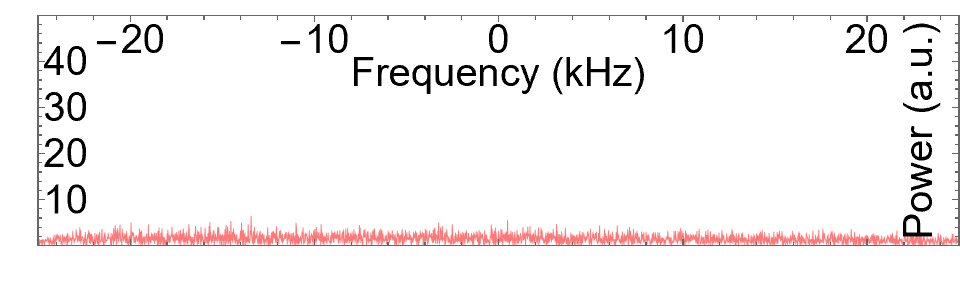} 
		& 1.3$\times$ \\
		& \makecell{Inner floating,\\ gradients connected,\\ GPA off} & \includegraphics[width=\figsize cm]{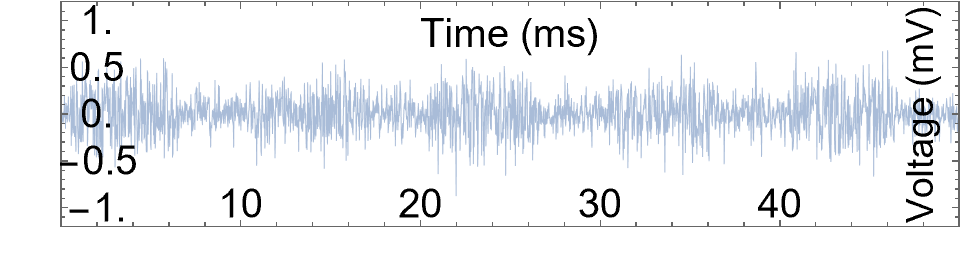} 
		& \includegraphics[width=\figsize cm]{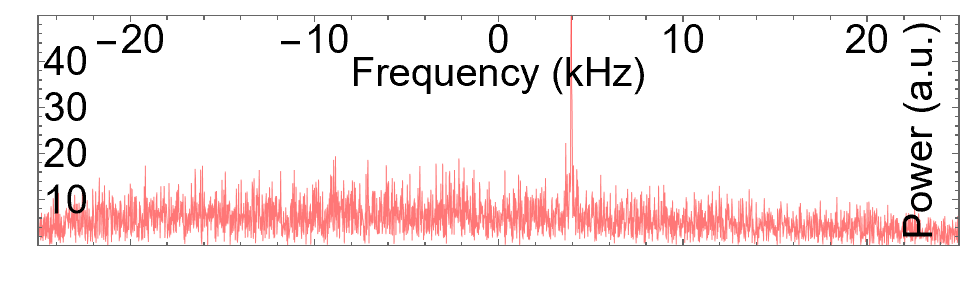} 
		& 5.3$\times$ \\
		& \makecell{Inner floating,\\ gradients connected,\\ GPA on} & \includegraphics[width=\figsize cm]{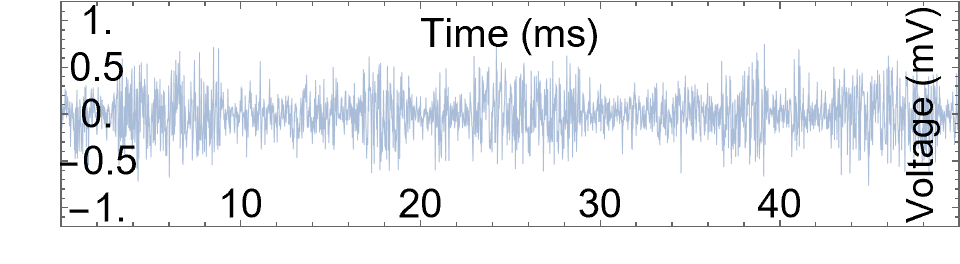} 
		& \includegraphics[width=\figsize cm]{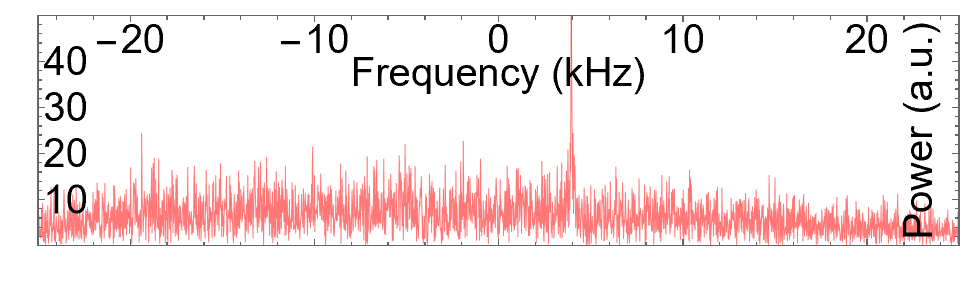} 
		& 5.9$\times$ \\
		& \makecell{Inner grounded,\\ external floating} & \includegraphics[width=\figsize cm]{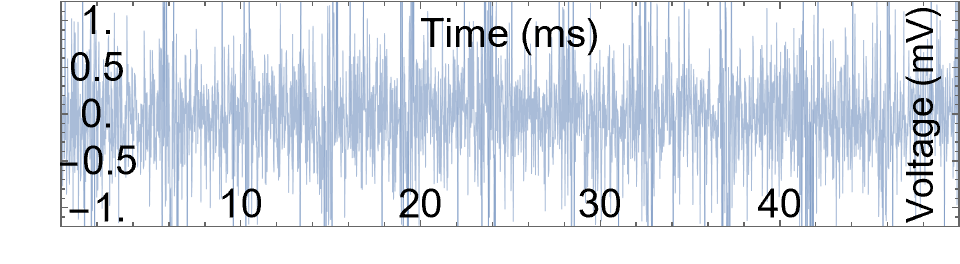} 
		& \includegraphics[width=\figsize cm]{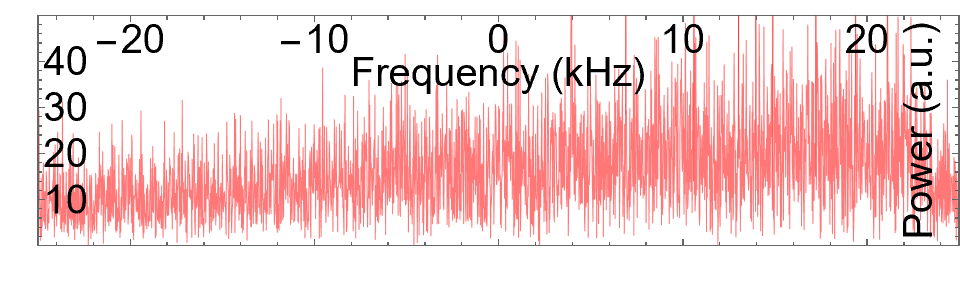} 
		& 14.8$\times$ \\
		
		\midrule
		
		\multirow{3}{*}{\makecell{Noise sources \\ (switched-mode \\ power supplies)}} 
		& \makecell{1\,m from\\ gradient cables} & \includegraphics[width=\figsize cm]{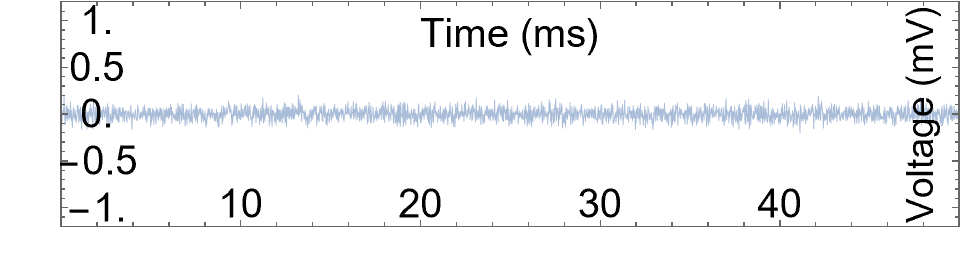} 
		& \includegraphics[width=\figsize cm]{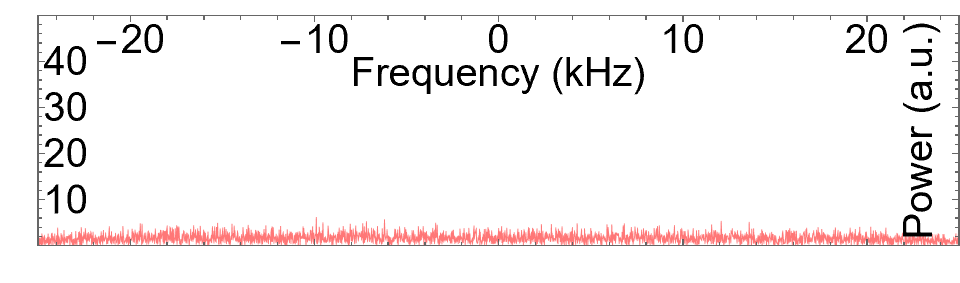} 
		& 1.5$\times$ \\
		& \makecell{10\,cm from\\ gradient cables} & \includegraphics[width=\figsize cm]{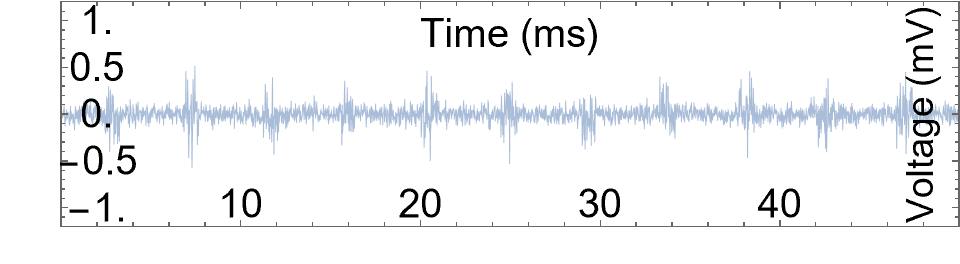} 
		& \includegraphics[width=\figsize cm]{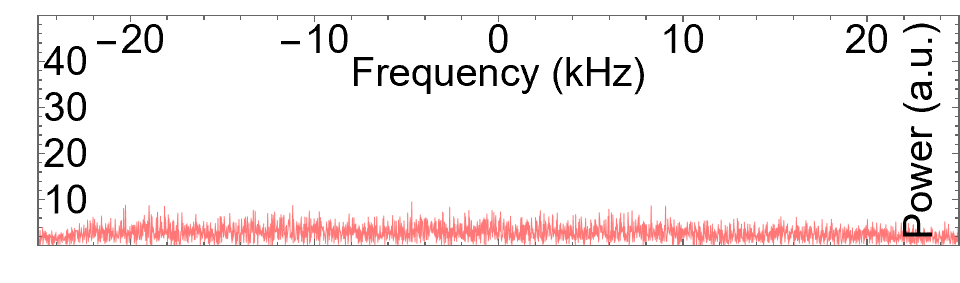} 
		& 2.4$\times$ \\
		& \makecell{In contact with\\ gradient cables} & \includegraphics[width=\figsize cm]{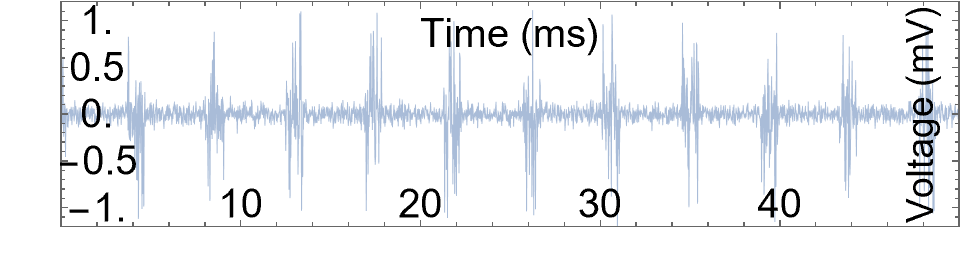} 
		& \includegraphics[width=\figsize cm]{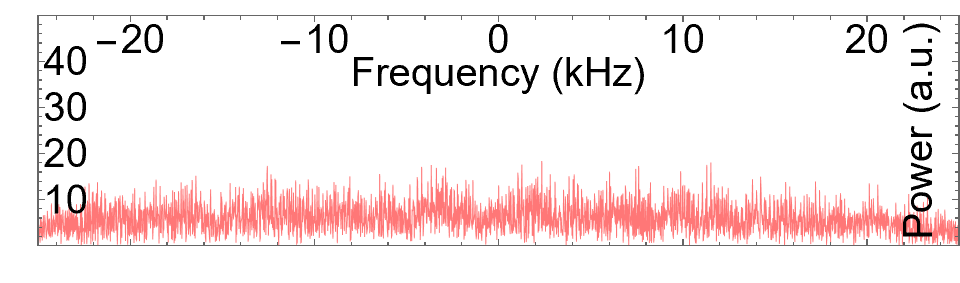} 
		& 5.4$\times$ \\
		
		\bottomrule
	\end{tabular}
\end{table*}

In addition to the protocol-driven noise characterization, we conducted targeted experiments to evaluate the effectiveness of selected wiring and shielding strategies outlined in the Appendix. Table~\ref{tab:time_trace_spectra} summarizes noise measurements obtained under different hardware configurations. Each configuration corresponds to a controlled variation of a single parameter—such as the presence or absence of internal and external RF shields, or the distance to known sources of noise.

\begin{figure*}[htbp]
	\centering
	\includegraphics[width=1\textwidth]{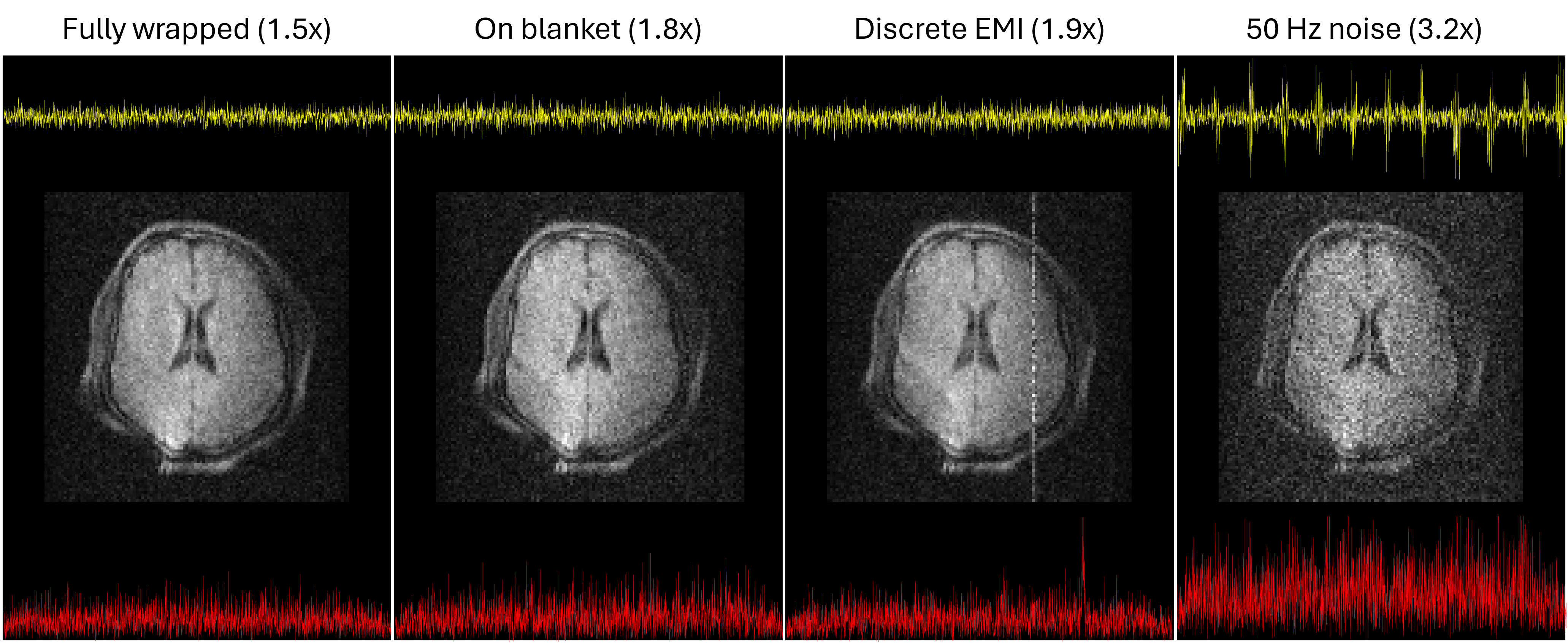}
	\caption{Comparison of \emph{in vivo} images acquired under different EMI suppression configurations using a grounded conductive blanket. From left to right: (a) blanket wrapped around the subject, (b) blanket under the subject, (c) blanket wrapped around the subject, with the MaRCoS and TM boxes open to couple a discrete EMI, and (d) blanket wrapped around the subject, with a switched-mode power supply next to the gradient lines to couple 50\,Hz noise. Noise measurements are shown also for reference. The top traces (yellow) correspond to the real time signals over the 50\,ms acquisition. The bottom traces (red) correspond to the power spectra over the 50\,kHz bandwidth. The axes scales are the same across all plots.}
	\label{fig:cloth_images}
\end{figure*}

Finally, to illustrate the practical consequences of subject-coupled EMI, Figure~\ref{fig:cloth_images} presents a series of \emph{in vivo} images acquired under different conditions. All scans were performed using the same imaging sequence and scanner setup. In the first condition, the subject was fully wrapped in the grounded cloth. In the second, the cloth was placed underneath the subject. For the third, we removed the lids from the MaRCoS and TM boxes to couple a discrete EMI spike into the receive chain (as in Table~\ref{tab:time_trace_spectra}). For the final image, we placed a switched-mode power supply unit in close proximity to the gradient cables to couple noise from the mains (similar to the last row in Table~\ref{tab:time_trace_spectra}).

\section{Discussion}

\subsection{Progressive Noise Characterization}

A seemingly unremarkable but central outcome of this work is summarized in Table~\ref{tab:noise_stages}, which presents measured noise levels throughout each stage of system assembly. This progressive characterization allowed us to bring a home-built low-field MRI scanner to a fully operational state with receive noise performance approaching the theoretical thermal limit.

Initial tests using a \SI{50}{\ohm} load established a clean baseline and verified the absence of significant excess noise in the core receive chain. Although components such as transmit electronics, gradient drivers, and auxiliary subsystems are not directly part of the receive path in this configuration, their physical presence can still affect system performance. Ground loops, shielding discontinuities, or poor cable layout can introduce EMI even when these components are idle. For this reason, noise measurements at each integration step remain essential.

With the RF coil installed and properly tuned and matched, additional care is needed. Components added in steps II through IV—including the Tx/Rx switch, transmit chain, and RF power amplifier—are not expected to introduce substantial excess noise, though minor increases are common due to electromagnetic coupling. These effects reflect the intrinsic sensitivity of MRI receive coils and are typically mitigated by adhering to the wiring and shielding guidelines outlined in the Appendix. At this stage, a noise level below approximately 1.7$\times$ the thermal baseline is considered acceptable, while levels near 1.3$\times$ indicate excellent isolation. Larger increases often signal issues with RFPA power supplies, particularly in custom-built setups that lack RF blanking.

The introduction of gradient cables and the gradient power amplifier—steps V and VI—is frequently the most challenging. These components are physically proximate to the RF coil and can couple strongly through self-resonant effects in the gradient structure~\cite{Ronen2020}. In such cases, an effective inner RF shield is indispensable. Even with adequate shielding, careful attention must be given to cable routing, filtering, and ground return paths to preserve noise performance. The final integration of auxiliary systems (step VII), such as auto-TM modules or digital control electronics, can also inject noise unless well-isolated from the receive path.

Once the full system is assembled, \emph{in vivo} testing evaluates real-world performance under loaded conditions. Table~\ref{tab:noise_stages} includes measurements acquired with a human subject inside the scanner. In systems that have been optimized according to the protocol, \emph{in vivo} noise levels typically remain below 2$\times$ the thermal reference. Even in electromagnetically challenging setting values around 3$\times$ are consistently achievable. This is the case in a low-field installation in Uganda~\cite{Obungoloch2023}, whose performance we have recently boosted with the protocol presented in Sec.~\ref{sec:protocol} \cite{GuallartNaval2025}. In contrast, unmitigated subject coupling can raise noise levels by more than two orders of magnitude. This often occurs when parts of the subject's body extend beyond the RF shield, acting as antennas that introduce ambient EMI \cite{Lena2025}.

To suppress subject-coupled interference, physical mitigation strategies have proven effective. Grounded conductive garments wrapped around the subject can dramatically reduce EMI pickup~\cite{GuallartNaval2022,Algarin2024}. Alternatively, directly connecting the subject to the scanner ground using ECG patches or resistive wrist straps can minimize common-mode voltages~\cite{Lena2025}. Both methods are broadly applicable and reliable across systems.

Beyond physical suppression, active noise cancellation techniques provide complementary approaches. These methods typically sense EMI through auxiliary channels—either via the ``MR-silent'' mode of an RF coil~\cite{Parsa2023} or external reference antennas~\cite{Srinivas2022,Zhao2024}—and subtract it from the received signal. While such strategies may reduce visible noise artifacts in reconstructions, they do not eliminate noise from the system itself.

\subsection{EMI Pathways and Shielding Mechanisms}

Table~\ref{tab:time_trace_spectra} complements these results by demonstrating the critical importance of grounding and shielding practices in specific scenarios. Each configuration highlights how seemingly small changes in layout or isolation can produce extreme differences in measured noise. Beyond the absolute RMS values, readers are encouraged to examine both the time-domain traces and spectral profiles, as distinct EMI mechanisms leave characteristic signatures.

For instance, removing the shield from a single RF cable leads to a clear increase in broadband noise in the time trace, along with the appearance of a sharp spectral spike—indicative of a discrete EMI source—which would manifest as a zipper artifact in a Cartesian reconstruction (see Fig.~\ref{fig:cloth_images}). In contrast, unshielding one gradient cable causes only a modest rise in overall noise but similarly introduces a narrow spectral feature.

The experiments involving metallic enclosures reveal that opening the TM box has limited effect, but doing so when the MaRCoS box is simultaneously open leads to strong coupling from an internal EMI source—consistent with a known switching supply inside MaRCoS—to reach the receive chain. This highlights the importance of complete and continuous shielding of all subsystems.

The tests involving scanner shields provide some of the clearest illustrations. The highest noise levels are observed when the external shield is disconnected, underscoring its critical role in attenuating EMI from the gradient subsystem. Notably, most of the interference appears to couple through the gradient coils themselves: when the gradient connections are physically removed, the inner RF shield can even be floated without significant noise increase. These results indicate that the outer shield primarily protects against external EMI reaching the gradient wiring, while the inner shield provides secondary isolation for the RF coil. Together, they form a complementary shielding strategy—but the outer shield proves essential for suppressing an often dominant coupling pathway.

The proximity experiments with a switched-mode power supply demonstrate the strong spatial dependence of EMI coupling. Moving the supply from 1\,m away to direct contact with gradient cables results in a monotonic increase in noise, reinforcing the practical importance of component placement and cable separation in real-world installations.

\subsection{Imaging Consequences of Noise Control}

Finally, Figure~\ref{fig:cloth_images} illustrates the direct connection between noise performance and image quality. These \emph{in vivo} reconstructions, acquired under varying EMI suppression conditions, visually reinforce the impact of rigorous noise control. Note that these are the first brain images acquired with this apparatus, enabled by the noise suppression strategy presented in this paper. We have used rather standard pulse sequence parameters for this purpose and we have decided not to correct the obvious distortions due to main field inhomogeneity \cite{Borreguero2025}, as this changes the structure of the noise in the final reconstructions.

\section{Conclusion}

We have presented a comprehensive methodology for the identification, characterization, and suppression of electromagnetic noise in low-field MRI systems. By incrementally assembling the scanner and monitoring noise at each step, our protocol provides a practical roadmap to achieving operation near the fundamental thermal noise limit. The approach is validated through quantitative measurements (Table~\ref{tab:noise_stages}), targeted EMI mitigation strategies (Table~\ref{tab:time_trace_spectra}), and illustrative \emph{in vivo} imaging (Figure~\ref{fig:cloth_images}), all of which underscore the importance of rigorous noise control in ensuring image quality.

Active noise cancellation techniques—particularly those leveraging AI and deep learning~\cite{Liu2021}—have recently demonstrated remarkable potential for suppressing EMI in low-field MRI, and can improve the apparent quality of reconstructed images. Yet, to date, no studies have demonstrated that these approaches maintain stable performance across changes in environment, system configuration, EMI source characteristics, or subject variability. In particular, AI models are strongly dependent on the characteristics of the data presented during training, and may fail when exposed to EMI patterns, signal features, or system conditions not represented in the training set. Moreover, these networks may apply uncontrolled regularization that compromises signal fidelity in ways that are difficult to detect. In contrast, physical EMI suppression via RF engineering offers a predictable and robust baseline. Reducing noise at the coil raises the fundamental SNR ceiling and can improve the effectiveness of any downstream method, whether physical, algorithmic, or hybrid. For reference, in our lab, we operate all our scanners systematically close to thermal-compatible noise levels. This holds under vastly varying conditions and even when subject to extreme noise sources (without the conductive cloth, in Ref.~\cite{Algarin2024} we measured ambient noise several hundreds of times above the thermal limit). 

More broadly, the low-field MRI community should probably strive to converge on a strategy for addressing EMI. This process involves three essential steps. First, we must recognize that the ultimate performance limit is set by thermal noise—a fact well understood in RF engineering but often underappreciated in low-field system design. Second, we must adopt objective metrics that quantify system noise in absolute terms and directly in signal space, enabling reproducible and unbiased comparisons across platforms and methods. Third, we must acknowledge that operation near the thermal limit is achievable in practice—particularly when following structured integration and suppression protocols like the one presented here.

\bibliographystyle{ieeetr}

\begin{thebibliography}{10}

\bibitem{Webb2023}
A.~Webb and J.~Obungoloch, ``{Five steps to make MRI scanners more affordable
to the world},'' {\em Nature 2023 615:7952}, vol.~615, pp.~391--393, mar
2023.

\bibitem{Algarin2024}
J.~M. Algarín, T.~Guallart-Naval, E.~Gastaldi-Orquín, R.~Bosch,
F.~Juan-Lloris, E.~Pallás, J.~P. Rigla, P.~Martínez, J.~Borreguero,
R.~Alamar, L.~Martí-Bonmatí, J.~M. Benlloch, F.~Galve, and J.~Alonso,
``{Portable MRI for major sporting events - a case study on the MotoGP World
Championship},'' {\em Portable MRI Journal}, vol.~1, p.~1, 2 2024.

\bibitem{Shellock2024}
F.~G. Shellock, M.~S. Rosen, A.~Webb, W.~T. Kimberly, S.~Rajan, A.~N. Nacev,
and J.~V. Crues, ``{Managing Patients With Unlabeled Passive Implants on MR
Systems Operating Below 1.5 T},'' {\em Journal of Magnetic Resonance
Imaging}, vol.~59, no.~5, pp.~1514--1522, 2024.

\bibitem{Huang2025}
S.~Huang, J.~M. Algarín, J.~Alonso, R.~Anieyrudh, J.~Borreguero, F.~Bschorr,
P.~Cassidy, W.~M. Choo, D.~Corcos, T.~Guallart-Naval, H.~J. Han, K.~C. Igwe,
J.~Kang, J.~Li, S.~Littin, J.~Liu, G.~G. Rodriguez, E.~Solomon, L.-K. Tan,
R.~Tian, A.~Webb, S.~Weber, D.~Xiao, M.~Xu, W.~Yu, Z.~Zhang, I.~Zinghini, and
B.~Blümich, ``{Experience of how to build an MRI machine from scratch},''
{\em Progress in Nuclear Magnetic Resonance Spectroscopy}, vol.~150-151,
p.~101574, 2025.

\bibitem{Webb2023b}
A.~Webb and T.~O’Reilly, ``{Tackling SNR at low-field: a review of hardware
approaches for point-of-care systems},'' {\em Magnetic Resonance Materials in
Physics, Biology and Medicine}, 5 2023.

\bibitem{Ayde2024}
R.~Ayde, M.~Vornehm, Y.~Zhao, F.~Knoll, E.~X. Wu, and M.~Sarracanie, ``{MRI at
low field: A review of software solutions for improving SNR},'' {\em NMR in
Biomedicine}, vol.~38, no.~1, p.~e5268, 2025.

\bibitem{Lena2025}
B.~Lena, B.~de~Vos, T.~Guallart-Naval, J.~Parsa, P.~García-Cristóbal,
R.~van~den Broek, C.~Najac, J.~Alonso, and A.~Webb, ``{Subject grounding to
reduce electromagnetic interference for MRI scanners operating in unshielded
environments},'' {\em arXiv:2507.07459}, 2025.

\bibitem{Liu2021}
Y.~Liu, A.~T.~L. Leong, Y.~Zhao, L.~Xiao, H.~K.~F. Mak, A.~Chun, O.~Tsang,
G.~K.~K. Lau, G.~K.~K. Leung, E.~X. Wu, and X.~Linfang, ``{A low-cost and
shielding-free ultra-low-field brain MRI scanner},'' {\em Nature
Communications 2021 12:1}, vol.~12, pp.~1--14, dec 2021.

\bibitem{Srinivas2022}
S.~A. Srinivas, S.~F. Cauley, J.~P. Stockmann, C.~R. Sappo, C.~E. Vaughn, L.~L.
Wald, W.~A. Grissom, and C.~Z. Cooley, ``{External Dynamic InTerference
Estimation and Removal (EDITER) for low field MRI},'' {\em Magnetic Resonance
in Medicine}, vol.~87, pp.~614--628, 2 2022.

\bibitem{Parsa2023}
J.~Parsa, T.~O'Reilly, and A.~Webb, ``{A single-coil-based method for
electromagnetic interference reduction in point-of-care low field MRI
systems},'' {\em Journal of Magnetic Resonance}, vol.~346, p.~107355, 2023.

\bibitem{Zhao2024}
Y.~Zhao, L.~Xiao, J.~Hu, and E.~X. Wu, ``{Robust EMI elimination for RF
shielding-free MRI through deep learning direct MR signal prediction},'' {\em
Magnetic Resonance in Medicine}, 2024.

\bibitem{Johnson1928}
J.~B. Johnson, ``Thermal agitation of electricity in conductors,'' {\em
Physical Review}, vol.~32, no.~1, p.~97, 1928.

\bibitem{Galve2024}
F.~Galve, E.~Pallás, T.~Guallart-Naval, P.~García-Cristóbal, P.~Martínez,
J.~M. Algarín, J.~Borreguero, R.~Bosch, F.~Juan-Lloris, J.~M. Benlloch, and
J.~Alonso, ``{Elliptical Halbach magnet and gradient modules for low-field
portable magnetic resonance imaging},'' {\em NMR in Biomedicine}, p.~e5258, 3
2024.

\bibitem{Pallas2025}
E.~Pallás, P.~García-Cristóbal, F.~Galve, T.~Guallart-Naval,
M.~Fernández-García, J.~M. Algarín, J.~Borreguero, R.~Bosch, J.~Conejero,
J.~M. Benlloch, and J.~Alonso, ``{A low-field portable MRI scanner with an
elliptic-bore Halbach magnet},'' in {\em Proceedings of the 33rd Annual
Meeting of the International Society of Magnetic Resonance in Medicine},
2025.

\bibitem{Sarracanie2015}
M.~Sarracanie, C.~D. LaPierre, N.~Salameh, D.~E.~J. Waddington, T.~Witzel, and
M.~S. Rosen, ``{Low-Cost High-Performance MRI},'' {\em Scientific Reports},
vol.~5, p.~15177, dec 2015.

\bibitem{Negnevitsky2023}
V.~Negnevitsky, Y.~Vives-Gilabert, J.~M. Algar\'in, L.~Craven-Brightman,
R.~Pellicer-Guridi, T.~O'Reilly, J.~P. Stockmann, A.~Webb, J.~Alonso, and
B.~Menk\"uc, ``{MaRCoS, an open-source electronic control system for
low-field MRI},'' {\em Journal of Magnetic Resonance}, vol.~350, p.~107424,
2023.

\bibitem{GuallartNaval2022b}
T.~Guallart-Naval, T.~O'Reilly, J.~M. Algar\'in, R.~Pellicer-Guridi,
Y.~Vives-Gilabert, L.~Craven-Brightman, V.~Negnevitsky, B.~Menk\"uc,
F.~Galve, J.~P. Stockmann, A.~Webb, and J.~Alonso, ``Benchmarking the
performance of a low‐cost magnetic resonance control system at multiple
sites in the open marcos community,'' {\em NMR in Biomedicine}, vol.~36,
pp.~1--13, 2022.

\bibitem{Algarin2024b}
J.~M. Algarín, T.~Guallart-Naval, J.~Borreguero, F.~Galve, and J.~Alonso,
``{MaRGE: A graphical environment for MaRCoS},'' {\em Journal of Magnetic
Resonance}, vol.~361, p.~107662, 4 2024.

\bibitem{Layton2017}
K.~J. Layton, S.~Kroboth, F.~Jia, S.~Littin, H.~Yu, J.~Leupold, J.~F. Nielsen,
T.~St\"ocker, and M.~Zaitsev, ``Pulseq: A rapid and hardware-independent
pulse sequence prototyping framework,'' {\em Magnetic Resonance in Medicine},
vol.~77, pp.~1544--1552, 4 2017.

\bibitem{Bosch2024}
R.~Bosch, J.~M. Algar{\'\i}n, T.~Guallart-Naval, F.~Juan-Lloris, J.~Conejero,
and J.~Alonso, ``{An open-source automatic impedance tuning and matching
module for low-field systems in clinical settings.},'' in {\em Proceedings of
the 32nd Annual Meeting of the International Society of Magnetic Resonance in
Medicine}, 2024.

\bibitem{Ronen2020}
I.~Ronen, T.~O'Reilly, M.~Froeling, and A.~G. Webb, ``{Proton nuclear magnetic
resonance J-spectroscopy of phantoms containing brain metabolites on a
portable 0.05 T MRI scanner},'' {\em Journal of Magnetic Resonance},
vol.~320, p.~106834, 2020.

\bibitem{Obungoloch2023}
J.~Obungoloch, I.~Muhumuza, W.~Teeuwisse, J.~Harper, I.~Etoku, R.~Asiimwe,
P.~Tusiime, G.~Gombya, C.~Mugume, M.~H. Namutebi, M.~A. Nassejje,
M.~Nayebare, J.~M. Kavuma, B.~Bukyana, F.~Natukunda, P.~Ninsiima,
A.~Muwanguzi, P.~Omadi, M.~van Gijzen, S.~J. Schiff, A.~Webb, and
T.~O'Reilly, ``{On-site construction of a point-of-care low-field MRI system
in Africa},'' {\em NMR in Biomedicine}, vol.~36, p.~e4917, 7 2023.

\bibitem{GuallartNaval2025}
T.~Guallart-Naval and \emph{et al.}, ``{\emph{In vivo} imaging with a low-cost
MRI scanner and cloud data processing in low-resource settings},'' {\em
Publication in preparation}, 2025.

\bibitem{GuallartNaval2022}
T.~Guallart-Naval, J.~Algar{\'{i}}n, R.~Pellicer-Guridi, F.~Galve,
Y.~Vives-Gilabert, R.~Bosch, E.~Pall{\'{a}}s, J.~Gonz{\'{a}}lez, J.~Rigla,
P.~Mart{\'{i}}nez, F.~Lloris, J.~Borreguero, {\'{A}}.~Marcos-Perucho,
V.~Negnevitsky, L.~Mart{\'{i}}-Bonmat{\'{i}}, A.~R{\'{i}}os, J.~M. Benlloch,
and J.~Alonso, ``{Portable magnetic resonance imaging of patients indoors,
outdoors and at home},'' {\em Scientific Reports 2022 12:1}, vol.~12,
pp.~1--11, jul 2022.

\bibitem{Borreguero2025}
J.~Borreguero, F.~Galve, J.~M. Algarín, and J.~Alonso, ``Zero-echo-time
sequences in highly inhomogeneous fields,'' {\em Magnetic Resonance in
Medicine}, vol.~93, no.~3, pp.~1190--1204, 2025.

\bibitem{deVos2024}
B.~de~Vos, R.~Remis, and A.~Webb, ``{Segmented RF shield design to minimize
eddy currents for low-field Halbach MRI systems},'' {\em Journal of Magnetic
Resonance}, vol.~362, p.~107669, 5 2024.

\bibitem{Joffe2011}
E.~B. Joffe and K.-S. Lock, {\em Grounds for grounding: A circuit to system
handbook}.
\newblock John Wiley \& Sons, 2011.

\end{thebibliography}

\section*{Acknowledgements}
We thank Andrew Webb and Beatrice Lena (LUMC, Netherlands) for discussions and the CADs for the spiral head coil, and the members of the labs where we have carried out noise characterization and suppression activities for their help in the definition of this protocol: the MRILab in Valencia (Spain), PhysioMRI Tech (Spain), Mbarara University of Science and Technology (Uganda), Singapore University for Technology and Design (Singapore), and University of Cape Town (South Africa).

\section*{Author contributions}
Experiments performed by TGN and JA. Data analysis and figure preparation by JA and TGN. Comparison with the theoretical framework and protocol definition by all authors. Manuscript written by JA and TGN, and revised by all authors.

\section*{Funding}
This work was supported by the European Innovation Council under grant 101136407, the Ministerio de Ciencia e Innovación of Spain under grant PID2022-142719OB-C22, and the European Partnership on Metrology under grant 22HLT02 A4IM.

\section*{Conflicts of interest}
TGN consults for and JMA and JA are co-founders of PhysioMRI Tech S.L.  

\appendix
\section{Wiring and Grounding Recommendations}
\label{appendix:wiring}

Clean, carefully planned wiring and grounding are absolutely essential in low-field MRI systems. Electromagnetic interference (EMI) often enters the receive chain through flawed cable practices, poorly shielded electronics, or improper ground referencing. The following sections provide detailed and experience-based guidance to help avoid these pitfalls.

\section{General cabling}
\label{sec:general_cabling}

\begin{itemize}
	\item Cables are often \textbf{the} problem — and often an elusive one. Noise issues that seem random or untraceable frequently originate from overlooked cable faults.
	\item Handle cables with extreme care: do not stretch, twist, or kink them. Use strain reliefs and avoid placing them where they may be stepped on or pinched. RF cables (coaxial) are particularly delicate and should be protected at all times.
	\item Use high-quality cables from reliable vendors.
	\item Label and color-code all cables for clarity, maintenance, and debugging.
	\item Keep cables as short and thick as possible. This is especially critical for power, ground, and gradient cables, where low resistance reduces voltage drops and power dissipation.
	\item Minimize the number of cable transitions and interconnections. Each connector introduces impedance discontinuities and potential noise entry points.
	\item Avoid coiling cables, which creates inductive loops prone to antenna effects and ground loops.
\end{itemize}

\section{RF cables}
\label{sec:rf_cables}

\begin{itemize}
	\item Long coaxial cables can lead to resonant cavity effects at low-field RF wavelengths. Keep them short and use as few transitions as possible.
	\item Never use T-pieces. If monitoring or tapping is required, use directional couplers or RF power splitters.
	\item When RF cables connect to a Faraday enclosure, ensure a solid electrical connection between the cable connector and the metal casing.
	\item Consider using cable traps or baluns to suppress common-mode currents that may flow along the outer shield of coaxial cables.
\end{itemize}

\section{Non-RF cables}
\label{sec:nonrf_cables}

\begin{itemize}
	\item These cables are often unshielded and can pick up or radiate EMI. Use shielding sleeves and ensure proper grounding of power, gradient, and digital cables (see Sec.~\ref{sec:scanner_shielding}).
	\item Again, cable traps and baluns may be useful for suppressing unwanted currents.
	\item Be especially careful with gradient cables — they are a major source of noise in low-field systems.
\end{itemize}

\section{Metallic enclosures}
\label{sec:metal_enclosures}

\begin{itemize}
	\item Enclose all electronics in conductive casings (e.g., aluminum boxes). While commercial RF modules are usually well-shielded, homemade circuits (e.g., TxRx switches, LNAs, tuning/matching units) require close attention.
	\item Ensure strong electrical contact between all parts of the enclosure (e.g., screw down lids firmly, ensure contacts are through large surfaces, rather than point contacts).
	\item Ensure low-impedance ground contact between printed circuit boards and the inside of the containing boxes.
	\item Avoid unshielded openings or through-hole connections. Use proper panel connectors wherever possible.
\end{itemize}

\section{Scanner shieldings}
\label{sec:scanner_shielding}

\subsection{Internal shielding}
\begin{itemize}
	\item The system's ground reference is defined by the return line of the RF coil. To isolate this from gradient-induced noise, an inner shield is typically used — often a cylindrical copper sleeve.
	\item Ensure a robust connection (short and thick) between the coil return and the inner shield, ideally soldered or screwed at both ends.
	\item Inner shields are thin and prone to mechanical degradation. If noise suddenly increases, they should be among the first suspects.
	\item While thicker shields provide better RF isolation, they may introduce eddy current issues if not properly slotted or segmented \cite{deVos2024}.
\end{itemize}

\subsection{External shielding}
\begin{itemize}
	\item Ideally, the external shield fully encloses the scanner core (magnet and gradient system), leaving only the bore openings.
	\item If the shield is assembled from multiple parts (e.g., barrel, lids, baseplate), ensure robust electrical contact at all joints.
	\item Ensure a high-quality connection between internal and external shields. This allows the more mechanically robust external shield to serve as a grounding point for cable sleeves, Faraday cages, and other subsystems. Follow a star grounding pattern wherever possible \cite{Joffe2011}.
\end{itemize}

\section{Noise sources}
\label{sec:noise_sources}

\begin{itemize}
	\item Keep all “dirty” components as far as possible from the RF chain and gradient cables. This includes digital electronics, control computers, power supplies, and switching regulators.
	\item Route “dirty” cables (digital, high-power, etc.) away from the RF path and gradient cables.
	\item Avoid switch-mode power supplies whenever possible. Linear power supplies are strongly preferred, but should be placed far enough from the magnet to avoid $B_0$ 50 or \SI{60}{Hz} modulation if they contain magnetic cores in the transformer.
\end{itemize}

\section{Documentation}
\label{sec:documentation}

\begin{itemize}
	\item Maintain a detailed grounding and connection diagram, including as many system elements as possible — ideally all of them. Update the diagram whenever any change is made.
	\item Keep logs of all tests, results, and relevant observations. Fighting noise often involves trial, error, and time-dependent behavior. Historical notes are often invaluable when diagnosing persistent or recurring issues.
\end{itemize}

\end{document}